\documentclass[reprint,superscriptaddress,nofootinbib,amssymb,aps,prl,floatfix]{revtex4-2}

\usepackage{graphicx}% Include figure files
\usepackage{dcolumn}% Align table columns on decimal point
\usepackage{bm}% bold math
\usepackage{physics}
\usepackage{datetime}
\usepackage{hyperref}
\usepackage{tabularx}
\usepackage{amsmath}
\usepackage{float}
\usepackage{xcolor}

\usepackage[labelformat=simple,labelsep=space,justification=justified,singlelinecheck=false]{caption}

\usepackage{ragged2e}  % Add this in preamble

\usepackage[export]{adjustbox}  % to preserve aspect ratio

\begin{document}

\title{Spin decoherence dynamics of Er$^{3+}$ in CeO$_2$ film}
\author{Sagar Kumar Seth}
 \affiliation{Pritzker School of Molecular Engineering, University of Chicago, Chicago, Illinois 60637, USA}
\affiliation{Materials Science Division, Argonne National Laboratory, Lemont, Illinois 60439, USA}
\author{Jonah Nagura}
 \affiliation{Pritzker School of Molecular Engineering, University of Chicago, Chicago, Illinois 60637, USA}
 \author{Vrindaa Somjit}
\affiliation{Materials Science Division, Argonne National Laboratory, Lemont, Illinois 60439, USA}
 \author{Aneesh Bapat}
 \affiliation{Pritzker School of Molecular Engineering, University of Chicago, Chicago, Illinois 60637, USA}
\affiliation{Materials Science Division, Argonne National Laboratory, Lemont, Illinois 60439, USA}
\author{Xinhao Li}
\affiliation{Center for Nanoscale Materials, Argonne National Laboratory, Lemont, Illinois 60439, USA}
\author{Gregory D. Grant}
 \affiliation{Pritzker School of Molecular Engineering, University of Chicago, Chicago, Illinois 60637, USA}
 \affiliation{Materials Science Division, Argonne National Laboratory, Lemont, Illinois 60439, USA}
\author{Ignas Masiulionis}
 \affiliation{Pritzker School of Molecular Engineering, University of Chicago, Chicago, Illinois 60637, USA}
 \affiliation{Materials Science Division, Argonne National Laboratory, Lemont, Illinois 60439, USA}
\author{Xu Han}
 \affiliation{Center for Nanoscale Materials, Argonne National Laboratory, Lemont, Illinois 60439, USA}
 \affiliation{Pritzker School of Molecular Engineering, University of Chicago, Chicago, Illinois 60637, USA}
\author{F. Joseph Heremans}
 \affiliation{Materials Science Division, Argonne National Laboratory, Lemont, Illinois 60439, USA}
 \affiliation{Center for Molecular Engineering, Argonne National Laboratory, Lemont, Illinois 60439, USA}
 \affiliation{Pritzker School of Molecular Engineering, University of Chicago, Chicago, Illinois 60637, USA}
 \author{Giulia Galli}
 \affiliation{Pritzker School of Molecular Engineering, University of Chicago, Chicago, Illinois 60637, USA}
  \affiliation{Materials Science Division, Argonne National Laboratory, Lemont, Illinois 60439, USA}
 \affiliation{Department of Chemistry, University of Chicago, Chicago, Illinois 60637, USA}
\author{David D. Awschalom}
 \affiliation{Pritzker School of Molecular Engineering, University of Chicago, Chicago, Illinois 60637, USA}
 \affiliation{Materials Science Division, Argonne National Laboratory, Lemont, Illinois 60439, USA}
 \affiliation{Center for Molecular Engineering, Argonne National Laboratory, Lemont, Illinois 60439, USA}
 \affiliation{Department of Physics, University of Chicago, Chicago, Illinois 60637, USA}
 \author{Supratik Guha}
 \affiliation{Pritzker School of Molecular Engineering, University of Chicago, Chicago, Illinois 60637, USA}
 \affiliation{Materials Science Division, Argonne National Laboratory, Lemont, Illinois 60439, USA}
 \affiliation{Center for Molecular Engineering, Argonne National Laboratory, Lemont, Illinois 60439, USA}
\author{Jiefei Zhang}
 \email{jfzhang@anl.gov}
 \affiliation{Materials Science Division, Argonne National Laboratory, Lemont, Illinois 60439, USA}
 \affiliation{Center for Molecular Engineering, Argonne National Laboratory, Lemont, Illinois 60439, USA}
 
\date{\today}

\begin{abstract}
Developing telecom-compatible spin-photon interfaces is essential towards scalable quantum networks. Erbium ions (Er$^{3+}$) exhibit a unique combination of a telecom (1.5 $\mu$m) optical transition and an effective spin-$1/2$ ground state, but identifying a host that enables heterogeneous device integration while preserving long optical and spin coherence remains an open challenge. We explore a new platform of Er$^{3+}$:CeO$_2$ films on silicon, offering low nuclear spin density and the potential for on-chip integration. We demonstrate a 38.8 $\mu$s spin coherence, which can be extended to 176.4\nobreakspace $\mu$s with dynamical decoupling. Pairing experiments with cluster correlation expansion calculations, we identify spectral diffusion-induced Er$^{3+}$ spin flips as the dominant decoherence mechanism and provide pathways to millisecond-scale coherence.
\end{abstract}

\maketitle

Optically-addressable solid-state spins have been widely studied for applications in quantum networks \cite{Network_RevModPhys2023, Awschalom2021,Awschalom2018}. Erbium ions (Er$^{3+}$), offering optical transitions in the telecom C-band from the 4f-electron, have gained significant attention for the development of robust spin-photon interfaces \cite{Awschalom2018,GOLDNER20151}. The choice of material host plays a critical role in determining achievable Er$^{3+}$ electron-spin coherence times, absorption linewidths, and optical addressability. Erbium ions doped into bulk crystals such as CaWO$_4$, Y$_2$SiO$_5$, and YVO$_4$, have demonstrated milliseconds-long spin coherence times \cite{LeDantec2021CaWO4,Raha2020,Welinski2019,Xie2021}, enabling the demonstration of quantum memory in the optical domain \cite{Craiciu2019,Dajczgewand2014,OptialM_PhysRevLett2010}, in the microwave regime for ensemble-based microwave-to-optical conversion \cite{Xie2021,trasduction_PhysRevA}, and as microwave quantum memories \cite{Probst2015}. 

Despite the progress in hosting milliseconds-long spin coherence, optically addressing Er$^{3+}$ spin states with high fidelity requires integration with photonic waveguide cavities \cite{Sinclair_multiplexing, FaraonTransduction} and superconducting resonators \cite{FaraonTransduction}. This compensates for the inherently low oscillator strength of Er$^{3+}$ ions and alleviates the need for high dopant concentrations \cite{Craiciu2019, OptialM_PhysRevLett2010}, enabling the isolation of single ions for spin-photon entanglement \cite{Jeff_PhysRevX2025}. Bulk crystals pose limitations for device fabrication and integration. Therefore, the search for Er$^{3+}$ spins in a solid-state host that simultaneously provides the ease of integration and exhibits lifetime-limited optical and spin coherence remains an open challenge.

To this end, we have explored the new platform of erbium-doped cerium dioxide (Er$^{3+}$:CeO$_2$) film on silicon \cite{GregAPLMat,zhang2023opticalspincoherence}. The low nuclear spin concentration in CeO$_2$ (only 0.04\% $^{17}$O) indicates the potential for hosting spins with long coherence \cite{kanai2022generalized}. The platform shows promising narrow optical homogeneous linewidth \cite{zhang2023opticalspincoherence}. These, combined with its compatibility with silicon and controlled synthesis, make Er$^{3+}$:CeO$_2$ a compelling candidate for building opto-electrical quantum devices.

In this Letter, we study the decoherence dynamics of Er$^{3+}$ spins in this platform and demonstrate a 38.8 $\mu$s spin coherence time at $\sim$ 77 mK. The dominant decoherence process is spectral diffusion-induced Er$^{3+}$ spin flips, primarily from Er-Er dipolar interactions, a strong process in CeO$_2$ due to its cubic crystal field symmetry. This finding is corroborated by both experimental observations and cluster correlation expansion (CCE) calculations. We also discuss possibility of further spectral diffusion suppression at lower temperatures and the effects of impurities, providing a route to enable millisecond-scale spin coherence times.

\begin{figure}
    \centering
    \includegraphics{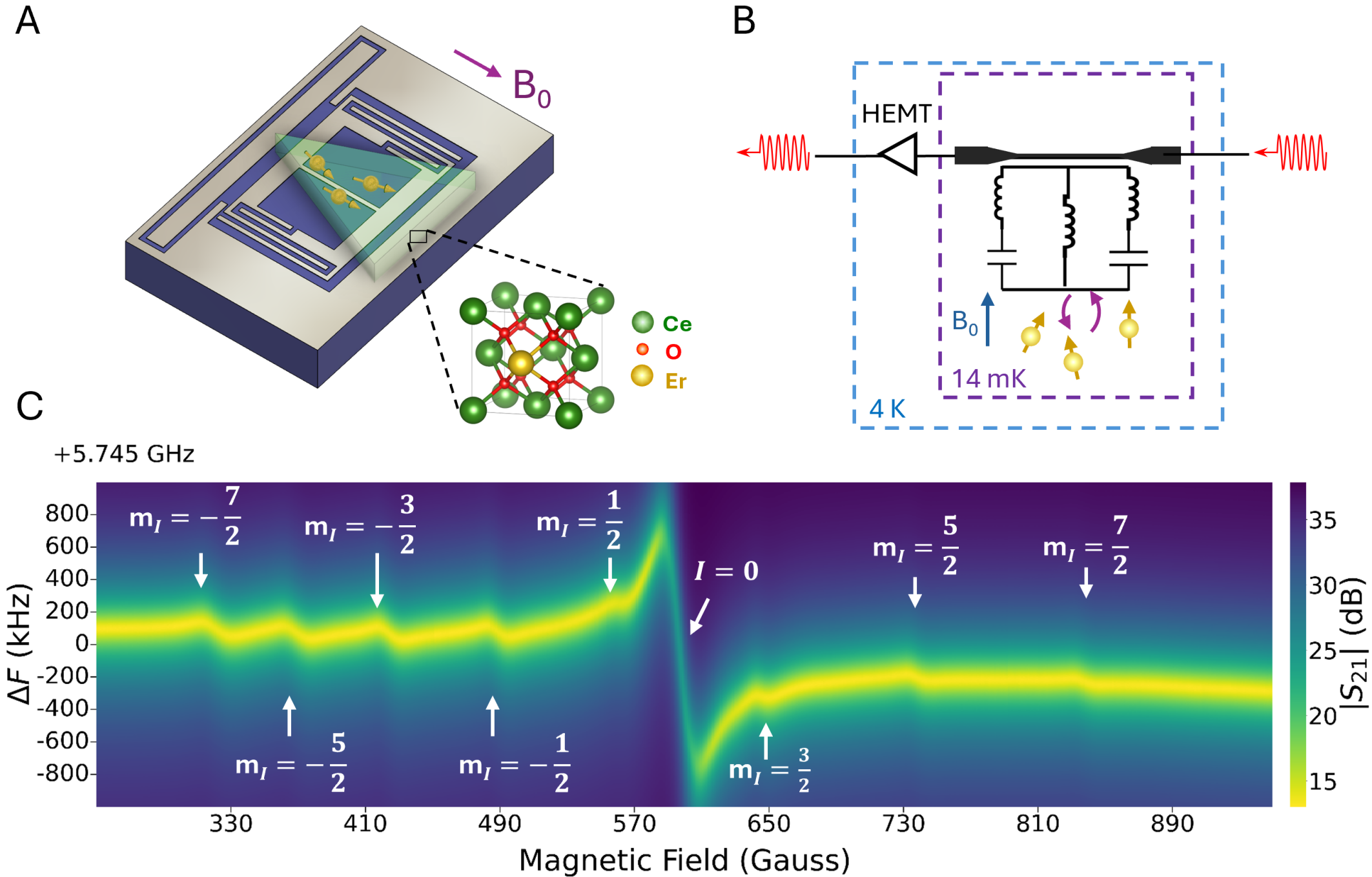}
    \caption{\justifying ESR spectroscopy. (a) Schematic of the coupling of Er$^{3+}$ spins (yellow) in CeO$_2$ film (glassy green) with the niobium (gray) superconducting LC resonator on sapphire (dark blue). (b) Schematic of experimental setup. (c) CW ESR spectrum measured with T$_{\text{MXC}}$=14 mK.}
    \label{fig:ESR}
\end{figure}

The sample used is a single-crystal CeO$_2$ film, \textit{in situ} doped with 2 ppm natural abundant Er$^{3+}$ ions, grown epitaxially on silicon. Fig.~\ref{fig:ESR}(a) and (b) show schematic descriptions of the sample coupled to superconducting resonator and measurement setup, respectively. The resonator has a fundamental frequency of $\omega_r/2\pi$=5.74 GHz and an internal quality factor greater than 1 $\times$ 10$^{5}$ (supplementary information (SI), SI. S1), sufficient for high-sensitivity spin detection near the inductor. Fig.~\ref{fig:ESR}(c) shows a CW (continuous wave) ESR spectrum of Er$^{3+}$ isotopes in the CeO$_2$ film as a function of the DC magnetic field B$_0$ applied parallel to the inductor wire. 

The largest anticrossing at 589 G is from $I$ =\nobreakspace 0 Er$^{3+}$ isotopes, predominantly $^{166}$Er$^{3+}$, $^{168}$Er$^{3+}$, and $^{170}$Er$^{3+}$. The ESR-allowed $\Delta$m$_I$ = 0 transitions between $^{167}$Er$^{3+}$ hyperfine levels are annotated according to their nuclear-spin projection $|m_I \rangle$. The positions of the observed anticrossings indicate a g-factor of 6.81$\pm$ 0.01, consistent with our previous studies \cite{GregAPLMat,zhang2023opticalspincoherence}. By analyzing the coupling strengths of various hyperfine transitions, we estimate the actual sample temperature to be T$_{\text{bath}}$=77.2 $\pm$ 13.5 mK (SI. S2). This slightly elevated temperature plays a key role in affecting the spin decoherence dynamics, as discussed later. 

We focus on the $|m_I \rangle  = |-1/2 \rangle$ spin sub-ensemble of $^{167}$Er$^{3+}$. This spin sub-ensemble couples to the resonator with an estimated ensemble coupling $g_{\text{ens}}/2\pi$ of 1.87 $\pm$ 0.05 MHz (SI. S3). This indicates a total number of 4.71 $\pm$ 0.26 $\times$ 10$^{5}$ spins coupled to the resonator \cite{KuboPRL2010, gavgPRL2017} (SI. S3), consistent with the 2 ppm Er$^{3+}$ concentration estimated from growth parameters. 

\begin{figure*}
    \centering
    \includegraphics{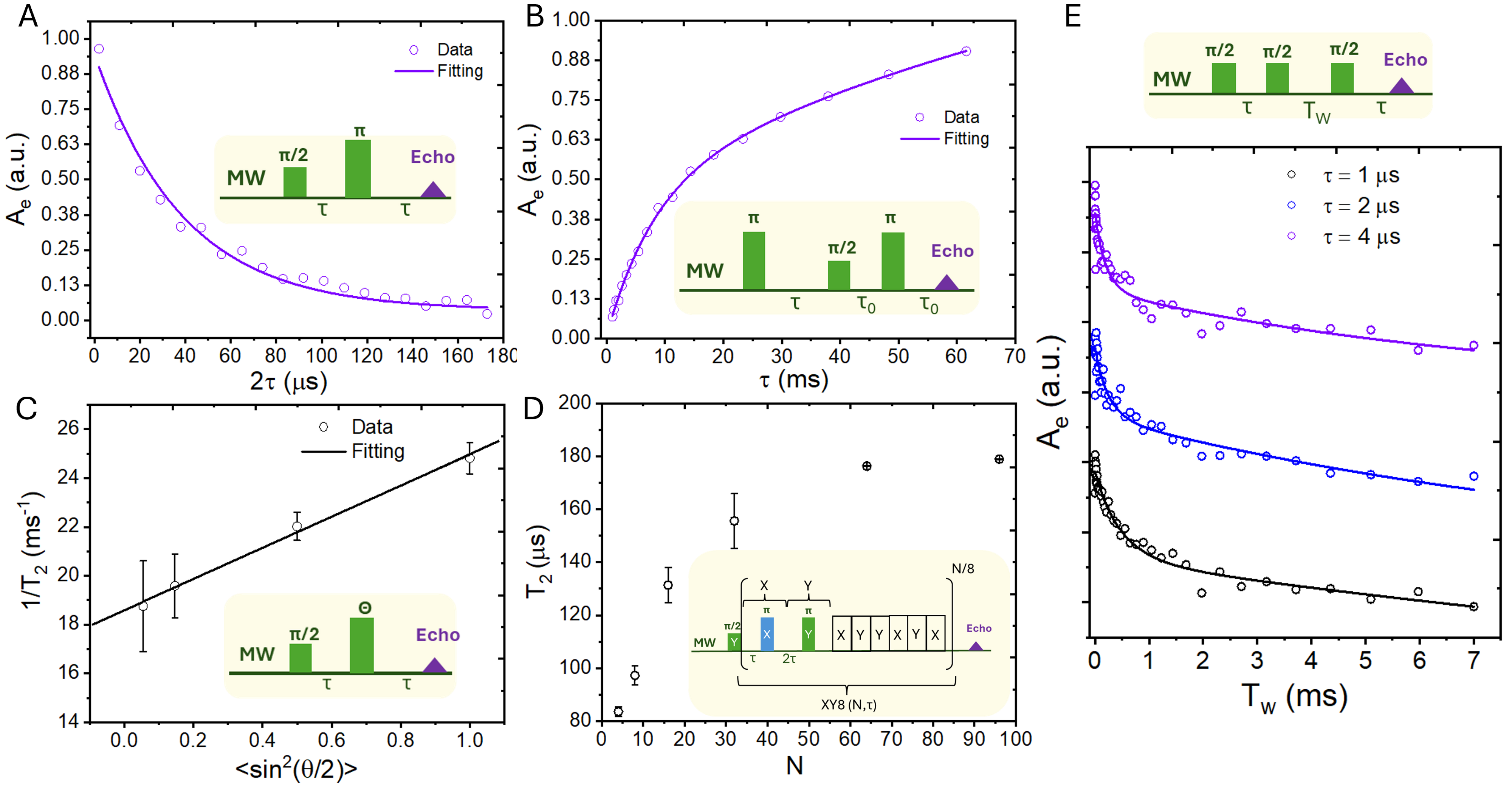}
    \caption{\justifying  Spin coherence and relaxation (a) Measured normalized spin echo integral, A$_e$, as a function of 2$\tau$. (b) Spin echo intensity measured with a three-pulse population inversion sequence as a function of $\tau$. (c) Plot of the inverse of T$_2$, extracted from the fits to the generalized Hahn-echo data as a function of averaged inversion pulse fidelity $<{\mathrm{sin}}^2(\theta/2)>$. (d) Plot of T$_2$ extracted from the XY8(N,$\tau$) measurements as a function of total number of pulses, N. (e) Plot of three-pulse echo decay as a function of T$_W$ at different $\tau$ values. The solid lines in the panels are fits to the data.} 
    \label{fig:T12}
\end{figure*}

We use a two-pulse Hahn-echo sequence to characterize the spin coherence, T$_2$, of this $^{167}$Er$^{3+}$ $|m_I \rangle  = |-1/2\rangle$ spin transition. The decay of the measured normalized spin echo integral, A$_e$, shown in Fig.~\ref{fig:T12}(a), indicates a spin T$_2$ of 38.8 $\pm$ 1.8 $\mu$s at sample temperature T$_{\text{bath}}$=77.2 $\pm$ 13.5 mK. We also examine the spin relaxation time, T$_1$, to probe spin-phonon interactions. The data (Fig.~\ref{fig:T12}(b)) suggests two relaxation pathways with a long T$_1$ of 119.3 $\pm$ 1.6 ms and a short T$_1$ of 8.0 $\pm$ 1.5 ms. The observed T$_2$ is orders of magnitude smaller compared to T$_1$, indicating that spin-phonon interactions are not the dominant mechanism for spin flip-flops.

We probe the contribution of instantaneous diffusion from spin-excitation-induced Er-Er spin flip-flops, along with the spectral diffusion induced by Er$^{3+}$ spin flips due to interaction with other paramagnetic spins. We first probe instantaneous diffusion using a generalized Hahn-echo sequence (Fig.~\ref{fig:T12}(c)). The instantaneous diffusion limited spin coherence, T$_{\text{2,ID}}$, is proportional to $\langle{\mathrm{sin}}^2(\theta/2)\rangle$ where $\theta$ is the rotation angle of the second pulse \cite{zhang2023opticalspincoherence,Salikhov1981}. With reducing $\theta \rightarrow$ 0, we find T$_2 \rightarrow$ 54.0 $\pm$ 0.7\nobreakspace $\mu$s. This indicates that the spectral diffusion limited T$_{\text{2,SD}}$ is 54.0 $\pm$ 0.7 $\mu$s and correspondingly the T$_{\text{2,ID}}$ is 158.9 $\pm$ 12.5 $\mu$s. 

We can extend the spin T$_2$ to 176.4 $\pm$ 0.1 $\mu$s using the XY8(N,$\tau$) sequence with 64 pulses (Fig.~\ref{fig:T12}(d)). The XY8 sequence does not symmetrize the Hamiltonian \cite{DD_PhysRevX2020}, and so it cannot cancel out the instantaneous diffusion while still potentially mitigating spectral diffusion \cite{DD_PhysRevX2020}. This observed limit on XY8 T$_2$ is consistent with the T$_{\text{2,ID}}$ extracted from instantaneous diffusion measurements (Fig.~\ref{fig:T12}(c)), confirming again that spectral diffusion is the dominant dephasing mechanism. 

We further probe the spectral diffusion induced Er$^{3+}$ spin flip rate using a three-pulse echo sequence \cite{SD1961PhysRev, YSO_montana_PhysRevB, CaWO4_SD_PhysRevB2022}. We study spin echo decay as a function of T$_W$ with five different $\tau$ values, ranging from 0.9 to 4 $\mu$s. Fig.~\ref{fig:T12}(e) shows three of the five measured three-pulse echo decay data. The echo integral, A$_e$, follows \cite{YSO_montana_PhysRevB, CaWO4_SD_PhysRevB2022}: 
\begin{equation}
\begin{split}
    A_e =A_0 \exp(-2\pi\Gamma_{\text{eff}}\tau) \exp(-T_W/T_1) \\ 
    \Gamma_{\text{eff}}=\Gamma_0+\frac{\Gamma_{\text{SD}}}{2}[R\tau+1-\exp(-RT_W)] 
    \label{Ae}
    \end{split}\quad,
\end{equation}
\noindent where $\Gamma_0$ is the decoherence processes occurring on a timescale faster than the measurements and instantaneous diffusion from spin excitation, $\Gamma_{\text{SD}}$ is the FWHM contribution to the dynamic distribution of transition frequencies within the sub-ensemble, and R is the average spin-flip rate. We assume that there is one dominant paramagnetic perturbing spin species, inducing spin flips via magnetic dipole interaction with Er$^{3+}$ spins.  

From the fits to the data using Eq.~\ref{Ae}, we obtain an average spin flip rate R=3.2 $\pm$ 0.5 ms$^{-1}$ and a spectral diffusion linewidth $\Gamma_{\text{SD}}$=82.8 $\pm$ 8.9 kHz at T$_{\text{bath}}$=77.2 $\pm$ 13.5 mK. The significantly higher spin flip rate R compared with the spin relaxation rate T$_1$$^{-1}$ further confirms that the spin-phonon induced spin flip contributes negligibly to spectral diffusion. This also sheds light on the observed spin relaxation pathways (Fig.~\ref{fig:T12}(b)): the shorter spin relaxation time ($\sim$ 8 ms) is likely dominated by spin-spin interactions, while the long spin relaxation ($\sim$ 119 ms) might arise from the spin-phonon interactions. 

Any paramagnetic species with a concentration comparable to or higher than the probed sub-ensemble can contribute significantly to spectral diffusion. We carry out temperature-dependent three-pulse echo measurements to probe $\Gamma_{\text{SD}}(T)$ and R$(T)$, to gain insight on the dominant paramagnetic perturbing spins. The linewidth $\Gamma_{\text{SD}}(T)$ is proportional to the number of spins that can undergo spin flips and thus follows the Boltzmann statistics \cite{YSO_montana_PhysRevB,CaWO4_SD_PhysRevB2022} with $\Gamma_{\text{SD}}(T)=\nobreakspace\Gamma_{\text{MAX}}\sech^2(\frac{g_{\text{s}}\mu_sB_0}{2k_BT})$, where $g_{\text{s}}$ is the effective gyromagnetic $g$-factor of the paramagnetic spins along the direction of the applied field. Similarly, the average spin flip rate R$(T)$ has combined effects from spin flip-flops and spin-phonon interactions resonant with the probed Er$^{3+}$ spin transition frequency \cite{YSO_montana_PhysRevB,CaWO4_SD_PhysRevB2022},
\begin{equation}
\begin{split}
  R(T)& = \alpha_{\mathrm{ff}}\frac{g_{\mathrm{s \perp}}{ }^4 n_{\text{s}}^2}{\Gamma_s}\sech^2\Big(\frac{g_{\text{s}}\mu_sB_0}{2k_BT}\Big)\\ 
    & \qquad + \alpha_{\mathrm{ph}} g_{\text{s}}^3 B_0^5\coth\Big(\frac{g_{\text{s}}\mu_sB_0}{2k_BT}\Big)
   \label{R_T}   
    \end{split}\quad,
\end{equation}
\noindent where the spin flip-flop rate is sensitive to $g_{s \perp}$, the $g$ tensor component of the paramagnetic spins perpendicular to the direction of the applied field \cite{flipflopPhysRevB}. Note the $n_{\text{s}}$ and $\Gamma_{\text{s}}$ represent the density and inhomogeneously broadened absorption linewidth of the paramagnetic spins, respectively. 

\begin{figure}[ht]
   \centering
    \includegraphics{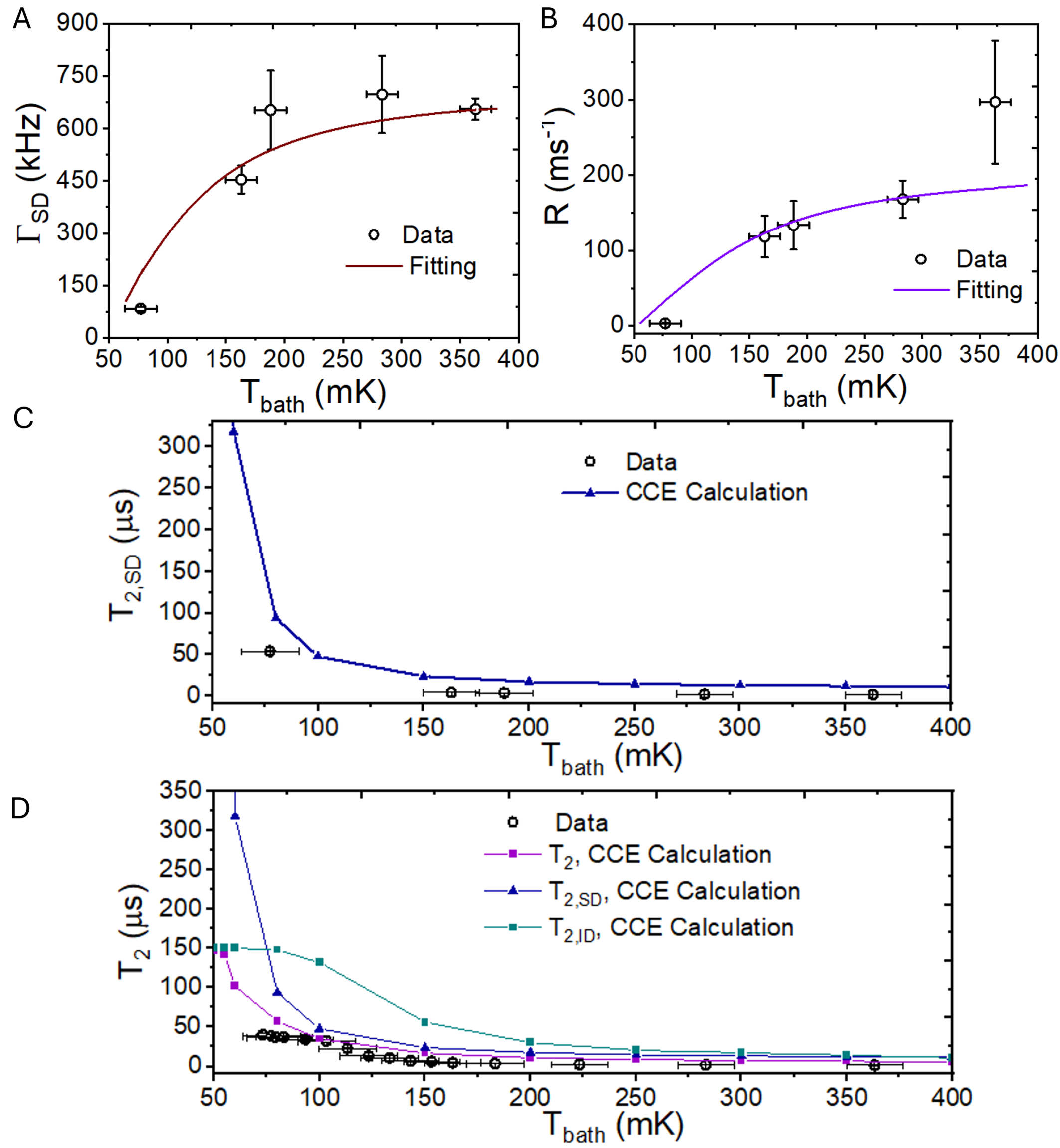}
    \caption{\justifying  Temperature-dependent spectral diffusion. (a) Linewidth $\Gamma_{\text{SD}}$ and (b) Spin-flip rate R extracted from three-pulse echo data. (c) Measured T$_{\text{2,SD}}$ and (d) spin T$_2$ as a function of sample temperature compared with CCE calculated results.}
    \label{fig:SD}
\end{figure}

Fig.~\ref{fig:SD}(a) shows the extracted temperature dependent spectral diffusion linewidth, $\Gamma_{\text{SD}}$(T). The solid line is a fit to the data, suggesting $\Gamma_{\text{MAX}}$ = 706.7 $\pm$ 48.5 kHz and $g_{\text{s}}$=6.17 $\pm$ 1.15. The spin-flip rate R(T) shows a similar trend (Fig.~\ref{fig:SD}(b)). The fitting (solid line, using Eq.~\ref{R_T}) results in an extracted value of g$_{\text{s}}$=7.34 $\pm$ 1.23. We have assumed $\alpha_{\mathrm{ph}}$=0 due to the order of magnitude higher spin-flip rate compared to the spin relaxation rate T$_1$$^{-1}$. Together, these data strongly suggest that spectral diffusion is predominantly driven by spin flips arising from dipolar interactions between Er$^{3+}$ ions, consistent with the high Er concentration in the sample. A similar effect has been observed in other bulk oxide materials \cite{CaWO4_SD_PhysRevB2022}. The higher spin-flip rate in our sample is likely due to the stronger Er–Er interactions, attributed to the larger transverse $g_{s \perp}$ ($\sim$ 6.8) due to the cubic crystal field symmetry, compared to other Er$^{3+}$-doped crystals with much smaller $g_{s \perp}$ ($<$ 2) \cite{CaWO4_SD_PhysRevB2022,V2o3_PhysRevApplied2023}. 

We leverage cluster correlation expansion (CCE) calculations \cite{onizhuk2024decoherencesolidstatespinqubits} to gain deeper insight into the spectral diffusion. Fig.~\ref{fig:SD}(c) shows the measured T$_{\text{2,SD}}$ as a function of temperature, T$_{\text{bath}}$, along with the calculated T$_{\text{2,SD}}$ using CCE, where the Er$^{3+}$ spins are treated as effective spin-1/2 systems with a g-factor of 6.8, randomly placed on the Ce$^{4+}$ sites of the CeO$_2$ lattice. The T$_{\text{2,SD}}$ is calculated only considering the spectral diffusion induced by the magnetic dipolar interactions between Er$^{3+}$ electron spins (SI. S6). The calculated temperature dependence of T$_{\text{2,SD}}$ agrees reasonably well with the experimental data, matching the expected trend. 

To develop a comprehensive understanding of the decoherence dynamics, CCE calculations on the T$_{\text{2,ID}}$ have also been carried out (SI. S6). The calculated results on T$_{\text{2,ID}}$ and T$_{\text{2,SD}}$ (Fig.~\ref{fig:SD}(d)) indicate: Spectral diffusion-induced decoherence is the dominant process at spin bath temperatures T$_{\text{bath}}$ $\geq$ 80 mK, while instantaneous diffusion dominates with T$_{\text{bath}}$ $\geq$ 400\nobreakspace mK. Considering both effects, the calculated T$_{\text{2}}$ is in reasonable agreement with the experimental data (black circles in Fig.~\ref{fig:SD}(d)), but with a slightly underestimated decoherence rate. The discrepancy could come from strong correlations and coupling between the bath spins due to the large Er$^{3+}$ g-factors. Future studies applying higher order or modified versions of CCE \cite{CCE_PhysRevB,CCE_PhysRevB2, MECCE_PRL} may better account for these correlations. The presence of paramagnetic impurities may also contribute to spectral diffusion. 

\begin{figure}[ht]
   \centering
    \includegraphics{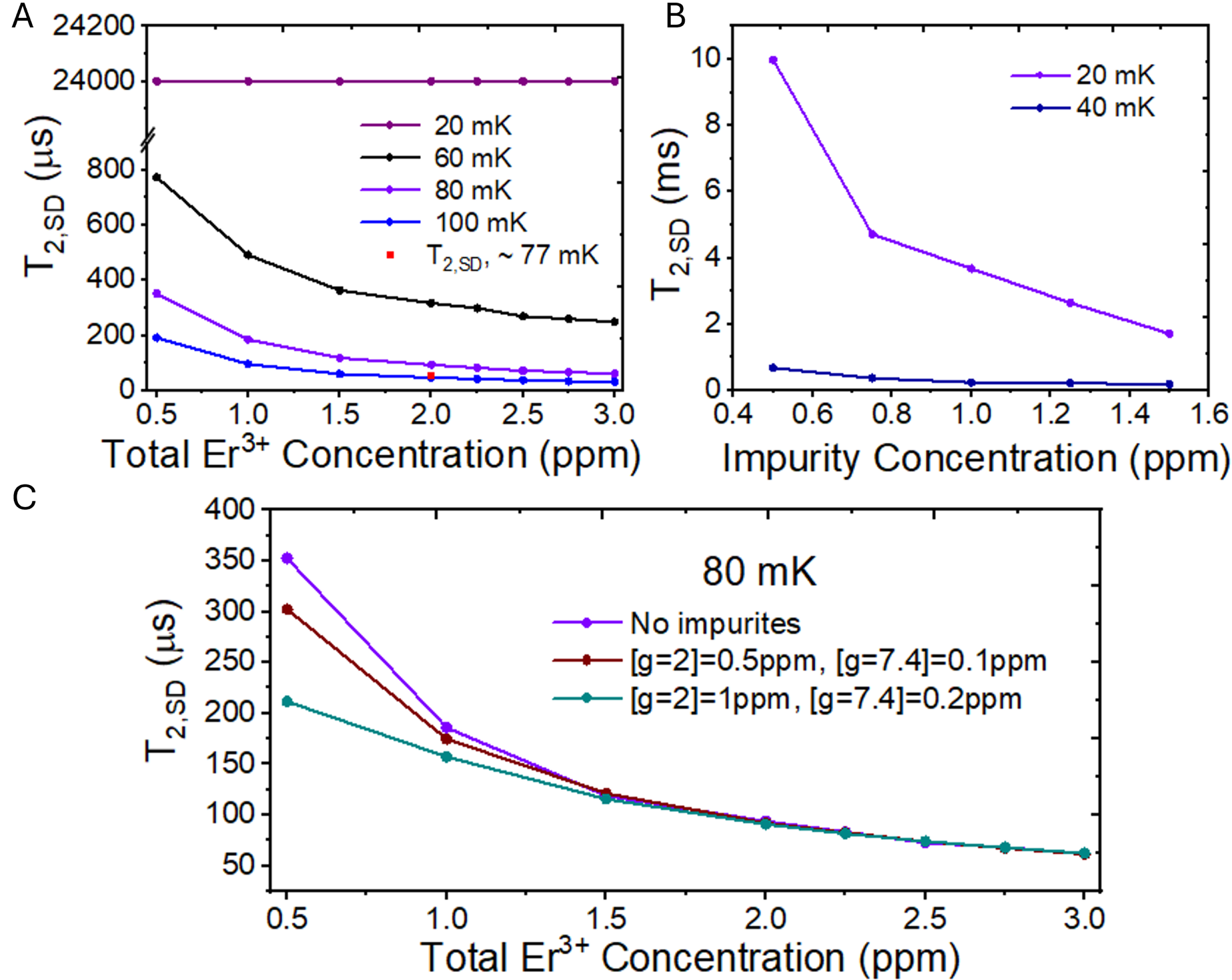}
    \caption{\justifying  CCE calculations of Er$^{3+}$ spin coherence. (a) Calculated T$_{\text{2,SD}}$ at 20-100 mK. (b) Calculated T$_{\text{2,SD}}$ accounting for the g=2 impurities and (c) both g=2 and g=7.4 impurities.}
    \label{fig:CCE}
\end{figure}

With spin bath temperature reduced down to 20 mK, the Er$^{3+}$ ensemble becomes fully polarized, quenching spectral diffusion from Er-Er interactions, leading to a predicted spin coherence time of 24 ms limited by interactions between Er$^{3+}$ electron spins and $^{17}O$ nuclear spins (Fig.~\ref{fig:CCE}(a)). At such low temperature, paramagnetic spins with smaller g-factors are not fully polarized and can induce Er spin flip-flops. One such known defect in CeO$_2$ is Ce$^{3+}$. Additionally, various impurities such as Gd$^{3+}$, Tb$^{3+}$, Cr$^{3+}$, and Dy$^{3+}$ can be incorporated during CeO$_2$ growth as evaporative contaminants from the cerium charge. Most impurities exhibit g $\sim$ 2 \cite{g_values_in_crystals}, except for Dy$^{3+}$ which has a g $\sim$ 7.4 \cite{g_values_in_crystals}. The presence of 0.5-1 ppm\nobreakspace g=2 impurities could reduce the T$_{\text{2,SD}}$ from 24 ms to 4 ms at 20 mK (Fig.~\ref{fig:CCE}(b)). The impact of g=2 impurities drastically reduces with increasing temperature. This indicates that these impurities do not strongly affect spectral diffusion in the temperature range explored in our study (Fig.~\ref{fig:CCE}(c)).

In our as-grown sample, the concentration of ions with g=2 is estimated to be $\sim$ 94.7$\pm$0.2 ppb (SI. S5). This signal arises from the cumulative contribution of Ce$^{3+}$ and other aforementioned g=2 paramagnetic impurities. Based on the growth conditions, we attribute this to Ce$^{3+}$. Dy$^{3+}$ transitions overlaps with the Er$^{3+}$ $|m_I \rangle  = |1/2 \rangle$ hyperfine peak, making it challenging to resolve. Nevertheless, as indicated by the CCE calculations, paramagnetic impurities could have a non-negligible impact on spin coherence at lower temperatures ($\leq$ 40~mK). A more accurate assessment of the concentration and spin relaxation dynamics of these impurities along with obtaining higher purity elements is essential for advancing our understanding of decoherence and guiding the development of high-coherence qubit systems. 

In conclusion, we examine the Er$^{3+}$: CeO$_2$ platform and report an electron-spin coherence time around 39\nobreakspace $\mu$s at millikelvin spin bath temperatures with nominal ppm doping. The dominant decoherence mechanism is identified as spectral diffusion, primarily driven by a high Er$^{3+}$ spin flip-flop rate, resulting largely from strong Er–Er magnetic dipolar interactions. Our results point to the feasibility of extending the coherence time into the millisecond regime with improved thermal anchoring and spin polarization of Er$^{3+}$ ions for quantum networks applications. The strong coupling between Er-Er in this platform could also enable study of correlated many-body spin dynamics with rich magnetic behaviors in long- and short-ranged interacting dipoles \cite{Dipolar_PhysRevResearch, spinexcahnge_PhysRevLett} and many-body entanglement \cite{douglas2024spinsqueezingmagneticdipoles}. 

\textit{Acknowledgments}\textemdash The authors would like to thank Dr. Shobhit Gupta and Dr. Jonathan Marcks for helpful discussions. This work is primarily funded by Q-NEXT, a U.S. Department of Energy Office of Science National Quantum Information Science Research Centers under Award Number DE-FOA-0002253, providing support for the sample growth, development of on-chip resonator and the spin measurements reported (S.K.S., A.B., G.D.G., I.M., S.G.). The spin measurements have additional support from the U.S. Department of Energy, Office of Science, Basic Energy Sciences, Materials Sciences and Engineering Division (F.J.H., D.D.A., J.Z.). The theory work (J.N., G.G.) is supported by MICCoM, which is part of the Computational Materials Sciences Program funded by the U.S. Department of Energy, Office of Science, Basic Energy Sciences, Materials Sciences, and Engineering Division. V.S acknowledges the support from the Maria Goeppert Mayer Named Fellowship, under the Laboratory Directed Research and Development (LDRD) funding from Argonne National Laboratory, provided by the Director, Office of Science, of the U.S. Department of Energy under Contract No. DE-AC02-06CH11357. This work made use of Pritzker Nanofabrication Facility, which receives support from the SHyNE; a node of the NSF’s National Nanotechnology Coordinated Infrastructure (NSF ECCS-1542205). Work performed at the Center for Nanoscale Materials, a U.S. Department of Energy Office of Science User Facility, is supported by the U.S. DOE, Office of Basic Energy Sciences, under Contract No. DE-AC02-06CH11357. 

\end{document}